\documentclass{article}
\usepackage{amsmath,amssymb}
\usepackage{graphicx}

\newcommand{\Hil}{\mathcal{H}}
\newcommand{\C}{\mathbb{C}}
\newcommand{\R}{\mathbb{R}}
\newcommand{\Sph}{\mathbb{S}}
\newcommand{\op}[1]{\hat{#1}}
\newcommand{\ket}[1]{|#1\rangle}
\newcommand{\bra}[1]{\langle #1|}
\newcommand{\brkt}[2]{\langle #1 | #2 \rangle}
\newcommand{\scprod}[2]{\text{\sf (} #1 \mathrel{,} #2 \text{\sf )}}
\newcommand{\Ps}{\op{\mathsf{P}}}
\newcommand{\Eq}[1]{Eq.~(\ref{#1})} 
\newcommand{\bbId}{{1\mkern -4.8mu\rm I}}
\newcommand{\Mat}[4]{\left#2\begin{array}{#1}#3\end{array}\right#4} 
 
\newcommand{\stack}[2][c]{\begin{array}{#1}#2\end{array}}
\newcommand{\ketup}{\ket{{\uparrow}}}
\newcommand{\ketdn}{\ket{{\downarrow}}}
\newcommand{\ie}{{\itshape i.e.},}
\newcommand{\Sec}[1]{Sec.~\ref{Sec:#1}} 
\newcommand{\Fig}[1]{Fig.~\ref{Fig:#1}} 

\newcommand{\ve}[1]{{\boldsymbol #1}}
\newcommand{\Id}{\op\bbId}
\newcommand{\TW}{\mathfrak{Z}}
\newcommand{\K}{\mathcal{K}}

\DeclareMathOperator{\Tr}{Tr}
\DeclareMathOperator{\RE}{Re}
\DeclareMathOperator{\IM}{Im}

\title{Classical simulators of quantum computers and no-go theorems}
\date{}
\author{Alexander Yu.\ Vlasov\thanks{%
Electronic mail: \protect\raisebox{-4pt}{\protect\includegraphics{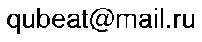}}}}
\begin{document}
\sloppy
\maketitle
\begin{abstract}
 It is discussed, why classical simulators of quantum computers 
escape from some no-go claims like Kochen-Specker, Bell, or recent
Conway-Kochen ``Free Will'' theorems.
\end{abstract}

\section{Introduction}

A term ``quantum computer'' was introduced just twenty five years ago, 
7 May 1981, by Feynman in his lecture ``Simulating Physics with Computers'' 
\cite{FeySim} at the conference on Physics and Computation.

In this lecture Feynman said \cite[p 471]{FeySim}
\begin{quote}
\small
[...]
``Might I say immediately, so that you know where I really
intend to go, that we always have had (secret, secret, close the doors!)
we always have had a great deal of difficulty in understanding the world
view that quantum mechanics represents. At least I do, because I'm an
old enough man that haven't got to the point that this stuff is obvious
to me. Okay, I still get nervous with it. And therefore, some of the younger
students {\ldots} you know how it always is, every new idea, it takes a
generation or two until it becomes obvious that there's no real problem.
It has not yet become obvious to me that there's no real problem. I cannot
define the real problem, therefore I suspect there's no real problem,
but I'm note sure there's no real problem. So that's why I like to
investigate things. Can I learn anything from asking this question
about computers---about this may or mat not be mystery as to what the
world view of quantum mechanics is?''
[...]
\end{quote}

He also explained, why ``asking questions about computers'' may be useful
\cite[p 486]{FeySim}
\begin{quote}
\small
[...]
``So, I would like to see if there's some other way out,
and I want to emphasize, or bring the question here, because
the discovery of computers and the thinking about computers 
has turned out to be extremely useful in many branches of
human reasoning. For instance, we never really understood
how lousy our understanding of language was, the theory of
grammar and all that stuff, until we tried to make a
computer which would be able to understand language. 
We tried to learn a great deal about psychology by trying to
understand how computers work. There are interesting 
philosophical questions about reasoning, and relationship,
observation, and measurement and so on, which computers
have stimulated us to think about anew, with new types of
thinking. And all I was doing was hoping that the computer-type of
thinking would give us some new ideas, if any are really needed.''
[...]
\end{quote}

\medskip

Being only embarked in theory of quantum computers about 15 years ago, 
author often encountered an idea that results like Kochen-Specker theorem
may prevent successful simulation of such quantum devices on classical 
computers. The doubts were dispelled later due to papers like \cite{Hol,GZ90} 
and appearance of first programs-simulators of quantum computers. 

In the Feynman lecture \cite{FeySim} as well as in later works, 
which provided base of modern theory of quantum computation 
\cite{DeuUQC,DeuQN,DeuJoz,Shor}, is written only 
about limitations due to exponential growth of complexity, but not 
about principle impossibility of modeling. 

Formally, any classical program-simulator is some {\em nonlocal} 
model of quantum computer and it let avoid limitation risen
by Bell consideration \cite{Bell}. On the other hand, it appears, 
that Kochen-Specker theorem \cite{KS,P} is concerned  with some 
other class of limitations.

It is interesting to apply ``computer-type of thinking''
by considering principles and problems of simulation of Bell 
experiments with two spin-half systems \cite{Bell}, model with 
spin-1 system used in Kochen-Specker theorem \cite{KS}, as well 
as experiments with both kinds of problems, like discussed in \cite{KC}.

\smallskip

In the \Sec{SimQC} is considered a minimalistic model, well known from 
theory of quantum computations and appropriate for consideration of most 
(thought) experiments with quantum systems used in no-go theorems mentioned 
above. It is applied first in \Sec{KS} to consideration of models used in 
Kochen-Specker theorem. Next sections are devoted to {\em nonlocal} models.
In \Sec{NlBell} is discussed {\em Bell pair}, {\ie} compound system with two 
qubits. Model with two spin-1 systems used in Conway-Kochen ``Free Will'' 
theorem is briefly analyzed in \Sec{3W}. In \Sec{DHT} is revised the problem
with relativity principle for measurement of nonlocal quantum systems. 

\section{Simulation of quantum computer}
\label{Sec:SimQC}

Minimal model with pure states may be defined as: 

\begin{flushleft}
\begin{itemize}
\item{\bf States:} Pure state is defined as normalized vector in Hilbert space
 \mbox{$\ket{\psi} \in \Hil = \C^N$}, $\brkt{\psi}{\psi}=1$.
(More precisely, the state is element of space of rays, {\ie}
 complex projective space\footnote{Complex projective space $\C P^{N-1}$
 is space of rays in $\C^N$, {\ie} factor space 
 $\mathbf{v} \sim \lambda \mathbf{v}$, $\mathbf{v} \in \C^N$,
 $\lambda \in \C$. If to express $\lambda = |\lambda| e^{i \theta}$, 
 it can be considered as space of normalized complex vectors (that is
 equivalent to real hyper-sphere $\Sph^{2N-1}$) modulo phase 
 (circle $\Sph^1$).}
  \mbox{$\C P^{N-1} \cong \Sph^{2N-1}/\Sph^1$}).
\item{\bf Composition:} State of compound system is described by tensor
 product of Hilbert spaces $\Hil_{12} = \Hil_1 \otimes \Hil_2 = \C^{N_1 N_2}$.
\item{\bf Gate:} Quantum gate is any unitary operator $U \in {\rm SU}(N)$ on 
 Hilbert space  of given quantum system $\ket{\psi'} = \op U \ket{\psi}$.
\item{\bf 1-Measurement (``filter''):} Such measurement is described by 
filter $F(\ket{\phi})$ and has analogue with question 
``Is it similar with $\ket{\phi}$?''  
that for some state $\ket{\psi}$ produces $\ket{\phi}$ (answer ``yes'') with 
probability $|\brkt{\psi}{\phi}|^2$ (square of scalar product) or ``nothing.''
\item{\bf N-Measurement (``basis''):} Measurement in given complete orthogonal 
basis $\ket{\phi_k}$, $k = 0,\ldots,N-1$. For given $\ket{\psi}$ produces 
one between $N$ outputs $\ket{\phi_k}$ with probability 
$p_k=|\brkt{\psi}{\phi_k}|^2$.
\item{\bf P-Measurement (``partial''):} It is given set of projectors
$\Ps_l$ corresponding to decomposition of $\Hil$ on direct sum of
orthogonal subspaces (not necessary rays). For given $\ket{\psi}$ produces 
with probability $p_l=\bra{\psi}\Ps_l\ket{\psi}$ 
normalized output $\ket{\phi_l} \equiv \Ps_l \ket{\psi}/\sqrt{p_l}$. 
\item{\bf $\boldsymbol{\op A}$-Measurement (values of operators)}. Let $\op A$ 
is (Hermitian) operator. It is used previous P-measurement procedure with 
$\Ps_l$ are projectors on eigenspaces of $\op A$. Now output with probability 
$p_l$ is corresponding eigenvalue, {\ie} $A_l=\bra{\phi_l}\op A \ket{\phi_l}$.
\end{itemize}
\end{flushleft}
\begin{quote}
\small
{\em References:} Chapters 1--3 of \cite{Dirac} for 
introduction to quantum theory and Dirac {\em bra, ket} notation,  
\cite{vnNeu} for von Neumann theory of quantum measurements, 
\cite{Min} for introduction to theory of quantum computation with 
pure states.
\end{quote}

Any classical simulator of quantum computer might be considered formally as
a hidden variable model of quantum system and so may be interesting for
general discussions on foundations of quantum mechanics. The model
discussed here is quite rough and can be simply adapted for realization
on classical computer: state is described as complex {\sf array} represented
as $2N$ real numbers, gate is a complex $N \times N$ matrix, 
measurements are reproduced by standard routines with random number generator.

The simple model could not substitute any ``serious'' interpretation
of quantum mechanics. It should be used with care, say measurements must
be only performed after all quantum gates and consequent measurements
are permitted only for commuting projectors
But such a model is enough for most examples in given paper. 

For more complicated schemes of quantum circuits, measurement and channels
may be used models with mixed states \cite{Mix}. 
Sometimes it is convenient to work with density matrix even
for pure state
\begin{equation}
  \op\rho = \ket{\psi}\bra{\psi}.
\label{rho}
\end{equation}
In such a case all formulas for probabilities used above should be rewritten
using equation like
\begin{equation}
 \bra{\psi}\Ps_l\ket{\psi} = \Tr(\Ps_l \op\rho).
\label{prho}
\end{equation}

It is also possible for some cases to consider even simpler model with real
spaces and spheres. Such kind of models are also illustrative, because very 
similar with some pure classical models used in theory of statistical
images (patterns) recognition \cite{Img}. Such property let us talk about 
such models not only as some reduced model of quantum mechanics, but also as 
about more or less ``natural'' example of classical statistical model and use 
it as test of some assumptions about hidden variables.

\section{Kochen-Specker theorem}
\label{Sec:KS}
\subsection{Simplified abstract model}
\label{Sec:KSimp}

Let us first consider simulation of quantum systems,
used in an abstract geometric version of Kochen-Specker model \cite{KS,P}, 
{\ie} it is considered sphere in {\em Euclidean} space $\R^N$, 
orthogonal frames (without relation with directions in physical space), 
and only {\em one-dimensional} projectors associated with 
vectors of the frames . 
Let us consider applications of model discussed in \Sec{SimQC} for 
three-dimensional case. It will be also used simplified version of
model from \Sec{SimQC} with simulation on real space 
instead of complex one, {\ie} sphere in 3D space.

A state is described by triple of real numbers $\ve{s}=(r,g,b)$, 
$r^2+g^2+b^2 = 1$, {\em aka} ``color (red, green, blue)''.
The measurements models here are very simple: {\bf 1-measurement} 
declares, that for filter, described by triple $\ve{f}=(r_f,g_f,b_f)$
the probability of ``yes'' answer is equal of square of scalar
product $p_c=\scprod{\ve{s}}{\ve{f}}^2$.  

An {\bf N-measurement} (here $N=3$) in a basis (frame) with three vectors 
$\ve{x}=(r_x,g_x,b_x)$, $\ve{y}=(r_y,g_y,b_y)$ and $\ve{z}=(r_z,g_z,b_z)$ 
is also quite straightforward: for any state $s$ and frame 
$(\ve{x},\ve{y},\ve{z})$ there are probabilities 

\begin{equation}
\begin{array}{c}
p_1 \equiv p_x=\scprod{\ve{s}}{\ve{x}}^2,\quad
p_2 \equiv p_y=\scprod{\ve{s}}{\ve{y}}^2,\quad 
p_3 \equiv p_z=\scprod{\ve{s}}{\ve{z}}^2,\\
p_x+p_y+p_z=1
\end{array}
\label{pxyz}
\end{equation}
that the measurement produces result $x$, $y$ or $z$ respectively.

\begin{figure}[hbt]
\begin{center}
\includegraphics[scale=0.5]{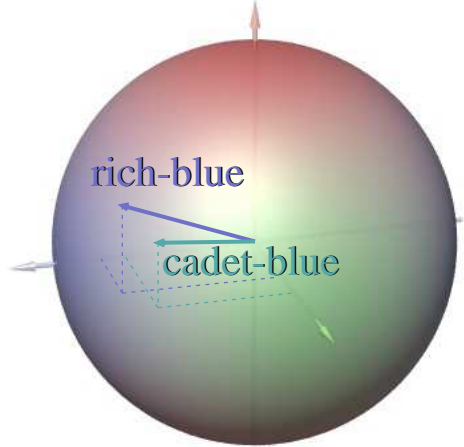}
\end{center}
\caption{Sphere of states (`colors') and vectors for two 
 different states}\label{Fig:blues}
\end{figure}

Formally, it is a hidden variables model, because here state is described 
by two real parameters (say two Euler angles on sphere), but any 
measurement may produce only discrete outcome with two (yes/no) or three values
(say 1, 2 or 3 for ``branches'' $\ve{x}$, $\ve{y}$ or $\ve{z}$ respectively).

It should be mentioned, that in such description the model is
not ``contextual,'' because probability of each outcome in 
\Eq{pxyz} depends only on given vector, but not other 
components of frame.


\smallskip

Let us use example with colors (\Fig{blues}). If we have some unknown color, then
for {\bf 1-measurement} it is possible to ask ``Is it blue color?'' 
If it really was some ``rich-blue color'' $\ve{c}_{rb}=(4/9,4/9,7/9)$, 
then ``yes'' has probability $(7/9)^2=49/81$ ($\sim 60.5\%$). It is possible 
also to use question with other sample, for example ``Is it cadet-blue color?'',
where ``cadet-blue color'' corresponds to vector $\ve{c}_{cb}=(1/3,2/3,2/3)$.
In such a case, answer ``yes'' has probability  
\mbox{$\scprod{\ve{c}_{rb}}{\ve{c}_{cb}}^2$}$ = (26/27)^2$ ($\sim 92.7\%$).

Let us consider now the {\bf 3-measurement} for $(r,g,b)$ frame. For
``rich-blue color'' it produces one of three outputs: ``red,'' ``green,''
or ``blue'' with probabilities 16/81, 16/81 and 49/81 respectively.
It should not be mixed with three independent questions, 
when it is possible to get from zero to three positive answers.
It is possible to calculate probabilities for zero, one, two and three 
answers ``yes'' respectively: 
\begin{eqnarray*}
P_{n=0}&=&(1-p_1)(1-p_2)(1-p_3),\\
P_{n=1}&=&p_1(1-p_2)(1-p_3)+(1-p_1)p_2(1-p_3)+(1-p_1)(1-p_2)p_3,\\
P_{n=2}&=&p_1 p_2 (1-p_3)+(1-p_1)p_2 p_3+p_1(1-p_2)p_3,\\ 
P_{n=3}&=&p_1 p_2 p_3. 
\end{eqnarray*}

For example with ``rich-blue'' color approximate values of the
probabilities are 25.4\% (0), 51.5\% (1), 20.7\% (2), 2.4\% (3),
{\ie} only for 51.5\% cases here is one ``yes''. 
Only if one of $p_k$ is unit and all other are zeros the
probability of one ``yes'' is 100\%.

\smallskip

\begin{figure}[hbt]
\begin{center}
\includegraphics[scale=0.5]{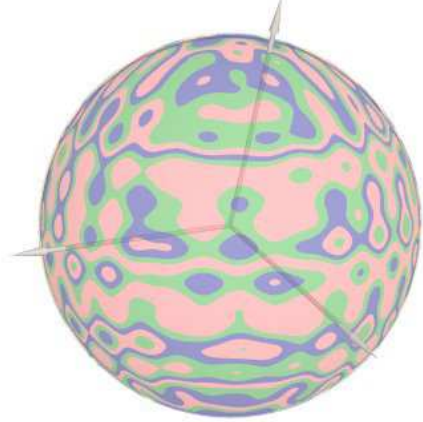}
\end{center}
\caption{Sphere for illustration of Kochen-Specker no-go theorem}\label{Fig:ks3c}
\end{figure}

Let us consider a geometric interpretation of Kochen-Specker theorem 
(\Fig{ks3c}): 
{\em It is impossible to have 3-coloring of sphere with property, 
that any three orthogonal arrows point to different colors.} 
It is equal reformulation of Kochen-Specker theorem with three colors 
instead of two numbers \cite{KS,P}, it is enough to assign unit to red 
and green areas and zero to blue ones. 

A short and understending formulation and interpretation of abstract 
Kochen-Specker theorem \cite[\S 7-3]{P}:
\begin{quote}
\small
``
The Kochen-Specker theorem asserts that, in a Hilbert space of dimension
$d \ge 3$, it is impossible to associate definite numerical values, 1 or 0,
with every projection operator ${\sf P}_m$, in such a way that, if a set of
{\em commuting} ${\sf P}_m$ satisfies $\sum {\sf P}_m = \bbId$, the corresponding 
values, namely $v({\sf P}_m)=0$ or $1$, also satisfy $\sum v({\sf P}_m) = 1$.
The thrust of this theorem is that any cryptodeterministic theory that
would attribute a definite result to each quantum measurement, and still
reproduce the statistical properties of quantum theory, is inevitably
{\em contextual}.
In the present case, if three operators, ${\sf P}_m$, ${\sf P}_r$, and ${\sf P}_s$,
have commutators $[{\sf P}_m,{\sf P}_r]=[{\sf P}_m,{\sf P}_s]=0$ and 
$[{\sf P}_r,{\sf P}_s] \ne 0$, the result of a measurement of ${\sf P}_m$ 
cannot be independent of whether ${\sf P}_m$ is measured alone, 
or together with ${\sf P}_r$, or together with ${\sf P}_s$.''
\end{quote}

\medskip

\Eq{pxyz} ensure that probability $p_x$ depends only on
$\ve{x}$ and coincides with probability for {\bf 1-measurement}, 
and so it does not matter, if instead of frame  
$(\ve{x}, \ve{y}, \ve{z})$ is used some other one 
$(\ve{x}, \ve{y}', \ve{z}')$ or $\ve{x}$ is measured alone. 
If the model of measurement
described in \Sec{SimQC} ``independent of whether  $\Ps_x$ 
is measured alone, or together with $\Ps_y$, or together with $\Ps_{y'}$''?

It is impossible to simulate measurements $\ve{x}$, $\ve{y}$, and $\ve{y}'$ 
in single run (here the model reproduces standard property 
of quantum mechanics) and so it is not possible to ask about 
definite results of such measurements --- they are not defined
in the model as well as in quantum mechanics. If to talk about
different runs, then even two measurements of the same $\ve{x}$
alone may produce different results. 

From such a point of view the model is not contextual, it is simply 
statistical, but for simulation of quantum system on deterministic 
classical computer it is necessary to suggest some algorithm for 
implementation of such statistical model. For generation of three
numbers with given probabilities $p_1$, $p_2$, $p_3$ 
$(p_1+p_2+p_3=1)$ it is commonly used generator of random numbers 
with uniform distribution $0 \le r \le 1$. The output 
$\{1,2,3\}$ is defined using inequalities: 
\begin{equation}
m(r,p_1,p_2,p_3)=
\begin{cases}
 1  : & 0 \le r < p_1,\\
 2  : & p_1 \le r < p_1+p_2, \\
 3  : & p_1+p_2 \le r \le p_1+p_2+p_3=1.
\end{cases}
\label{gen3}
\end{equation}
The method may be simply generalized for arbitrary number of outcomes
with rules for $k : \sum_{i=1}^{k-1} p_i \le p_k < \sum_{i=1}^k p_i$.

Here problem of contextuality may appear again:
 if to try to model different experiments
with same ``hidden variable'' $r$, then appearance of the value $2$ depends 
on $1$ via $p_1$ in \Eq{gen3}. 
But similar problems may appear in pure classical statistical ``paradoxes,''
for models with three outcomes. Well known example is
``paradox'' with division of a stick on three parts 
in classical theory of {\em geometrical probabilities} \cite{KM63,Gard}.
We may ask here questions similar with discussed above, {\ie} `What is 
hidden variable model for such division?', `How does length of first part is 
correlated with other ones?', or even, `Does it possible to suggest 
a model, that may ``naturally'' simulate this process?' The last question 
about existence of a model is not unusual, in fact, it is just a reason, 
why such a problem is called ``paradox.''

\begin{figure}[hbt]
\begin{center}
\includegraphics[scale=0.5]{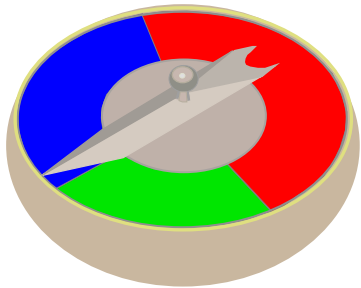}
\end{center}
\caption{Roulette with different sectors}\label{Fig:rul}
\end{figure}

It is possible to consider simpler example of pure classical statistical
``paradox,'' relevant with present discussion. Let we have roulettes with 
red, green and blue areas (with angles $2\pi p_r$, $2\pi p_g$, $2\pi p_b$)
for modeling an outcome with probabilities $p_1=p_r$, $p_2=p_g$, $p_3=p_b$
(\Fig{rul}). 
If appearance of green color depends only on value $p_g$? Yes, by definition! 
But in computer model of roulette with hidden parameter $r$ and algorithm 
like \Eq{gen3} it formally depends also on $p_r$, yet it may not be found 
by any statistical test.  

\subsection{Physical model}
\label{Sec:KSphys}

It is more interesting to consider simulation of model, used in
physical interpretation of Kochen-Specker theorem \cite{KS,P}. Here is
used presentation of momentum operators in symmetrical
form \cite{P} (system of units with $\hbar=1$)
\begin{equation}
 \op J_x = \Mat{ccc}({0&0&0\\0&0&-i\\0&i&0}) \!,
\quad
 \op J_y = \Mat{ccc}({0&0&i\\0&0&0\\-i&0&0}) \!,
\quad
 \op J_z = \Mat{ccc}({0&-i&0\\i&0&0\\0&0&0}) \!.
\label{JsP}
\end{equation}

Let us write eigenvectors of each operator:
\begin{subequations}\label{EJP}
\renewcommand{\theequation}{\theparentequation$_\alph{equation}$}
\addtocounter{equation}{23}
\begin{equation}
  \op J_x : \qquad 
  \ket{{+}_x} = \frac{1}{\sqrt{2}}\Mat{c}({0\\-i\\1}), \quad
  \ket{0_x} = \Mat{c}({1\\0\\0}), \quad
  \ket{{-}_x} = \frac{1}{\sqrt{2}}\Mat{c}({0\\i\\1}),
\label{EJxP}
\end{equation}
\begin{equation}
  \op J_y : \qquad 
  \ket{{+}_y} = \frac{1}{\sqrt{2}}\Mat{c}({-i\\0\\1}), \quad
  \ket{0_y} = \Mat{c}({0\\1\\0}), \quad
  \ket{{-}_y} = \frac{1}{\sqrt{2}}\Mat{c}({i\\0\\1}),
\label{EJyP}
\end{equation}
\begin{equation}
  \op J_z : \qquad 
  \ket{{+}_z} = \frac{1}{\sqrt{2}}\Mat{c}({-i\\1\\0}), \quad
  \ket{0_z} = \Mat{c}({0\\0\\1}), \quad
  \ket{{-}_z} = \frac{1}{\sqrt{2}}\Mat{c}({i\\1\\0}),
\label{EJzP}
\end{equation}
\end{subequations}
where eigenvectors $\ket{\pm_\nu},\ket{0_\nu}$, $\nu = x, y, z$ correspond to 
eigenvalues $\pm1, 0$ of operator $\op J_\nu$.

\smallskip

In Kochen-Specker theorem are used commuting operators \cite{KS,P}
\begin{equation}
 \op J_x^2 = \Mat{ccc}({0&0&0\\0&1&0\\0&0&1}) \!,
\quad
 \op J_y^2 = \Mat{ccc}({1&0&0\\0&0&0\\0&0&1}) \!,
\quad
 \op J_z^2 = \Mat{ccc}({1&0&0\\0&1&0\\0&0&0}) \!,
\label{JsqP}
\end{equation}
where $\op J_x^2 + \op J_y^2 + \op J_z^2 = 2$.
It is possible to measure all the values in one measurement of
an operator $a \op J_x^2 + b \op J_y^2 + c \op J_z^2$ \cite{KS},
{\em e.g.}, operator $\op\K$
\begin{equation}
 \op\K = 2\op J_x^2 + \op J_y^2 = \Mat{rrr}({1&0&0\\0&2&0\\0&0&3}).
\label{KN}
\end{equation}

The operator has eigenvalues $1,2,3$ with eigenvectors
\begin{equation}
 \op\K : \qquad 
  \ket{1_\K} = \Mat{c}({1\\0\\0}), \quad
  \ket{2_\K} = \Mat{c}({0\\1\\0}), \quad
  \ket{3_\K} = \Mat{c}({0\\0\\1}).
\label{EKN}
\end{equation}

It is possible to express $\op J_x^2, \op J_y^2, \op J_z^2$ using $\op\K$
\begin{equation}
  \op J_x^2 = \frac{1}{2}(4-\op\K)(\op\K-1), \quad
  \op J_y^2 = (\op\K - 2)^2, \quad
  \op J_z^2 = \frac{1}{2}(3-\op\K)\op\K.
\label{J2K}
\end{equation}

It is convenient, that $i \op J_\nu$ \Eq{JsP} in such presentation are real
matrixes and coincide with generators of SO$(3)$ group on real subspace
$\R^3$ of full Hilbert space $\C^3$ of spin-1 system. Eigenvectors
\Eq{EKN} of $\op\K$ also may be considered as elements of $\R^3$.
So geometrical model with real 3D space and its rotations (`$\op\K$-sphere')
used in Kochen-Specker theorem is justified here. 

Let us consider transition to some other frame, {\ie}
\begin{equation}
 \op\K' : \qquad 
  \ket{1_\K'} = \Mat{c}({x_1\\x_2\\x_3}), \quad
  \ket{2_\K'} = \Mat{c}({y_1\\y_2\\y_3}), \quad
  \ket{3_\K'} = \Mat{c}({z_1\\z_2\\z_3}).
\label{EK'}
\end{equation}
The transition \Eq{EK'} is described by some orthogonal
matrix $\op R$, composed from elements of the frame
\begin{equation}
 \op R = \Mat{ccc}({x_1&y_1&z_1\\x_2&y_2&z_2\\x_3&y_3&z_3}), \quad
 \op R^{-1}=\op R^{\sf T} = \Mat{ccc}({x_1&x_2&x_3\\y_1&y_2&y_3\\z_1&z_2&z_3}).
\label{Rot}
\end{equation}
It is simple to find expression for any $\op J_{\nu'}^2$ for new frame,
{\em e.g.}
\begin{equation}
\op J_{x'}^2 = \op R \op J_x^2 \op R^{-1} = 
 \Mat{ccc}({1-x_1^2&-x_1 x_2&-x_1 x_3\\
            -x_2 x_1&1-x_2^2&-x_2 x_3\\
            -x_3 x_1&-x_3 x_2&1-x_3^2}),  
\label{Jsqx'}
\end{equation}
and analogue expressions valid for $\op J_{y'}^2$ and $\op J_{z'}^2$.
For particular example of frame with same vector $\ve{x}$ the matrix is
\begin{equation}
 \op R_\theta 
 = \Mat{ccc}({1&0&0\\0&\cos(\theta)&\sin(\theta)\\0&-\sin(\theta)&\cos(\theta)})
\label{Rxy}
\end{equation}
and so $\op J_{x'}^2=\op R_\theta\op J_{x}^2\op R_\theta^{-1} = \op J_{x}^2$ 
(and $\op J_{x'}=\op R_\theta\op J_{x}\op R_\theta^{-1} = \op J_{x}$ in \Eq{JsP},
but $\op R_\theta\ket{\pm_x}$ in \Eq{EJxP} produces a phase multiplier).

It should be mentioned only, that geometrical interpretation with
directions in real physical space does not look so obvious. 
Eigenvectors of $\op\K$ coincide with eigenvectors $\ket{0_\nu}$ for
zero eigenvalues for three different operators $\op J_\nu$ \Eq{EJP},
but it is hardly to consider as some directions ``{\em zero} projections
of momentum''. Eigenvectors $\ket{\pm_\nu}$ for different momentum {\em are
not orthogonal} and have equal scalar products
$\brkt{J_x=1}{J_y=1}=\brkt{J_x=1}{J_z=1}=\brkt{J_z=1}{J_y=1}=1/2$
and so a visual classical picture with three orthogonal axes is not 
quite justified. 

The eigenvalue $1$ of $\op J_\nu^2$ corresponds to
2D eigenspace, {\ie} after normalization it is some subspace 
in $\C^3$ isomorphic to Riemann sphere, that intersects $\op\K$-sphere 
along big circle and so picture is rather complicated. 
The quantum mechanical description of spin-1 particle essentially 
differs from classical, it may not be presented as some vector in 3D 
space. In model with pure states from \Sec{SimQC} it is described by
complex projective space $\C P^2$, that may be represented as
nontrivial four-dimensional real manifold $\Sph^5/\Sph^1$, but not as
classical model with sphere $\Sph^2$ of unit vectors in $\R^3$. 

\smallskip

Let us now consider measurements of $\op J_\nu^2$ using model from
\Sec{SimQC}. 
If there is some state $(k_1,k_2,k_3)$, then measurement of $\op\K$ 
produces $1$, $2$, $3$ with probabilities $|k_1|^2$, $|k_2|^2$, $|k_3|^2$ 
respectively. Each value of $\K$ corresponds to certain combination of values 
$J_x^2$, $J_y^2$, $J_z^2$, {\ie} 
$1 \mapsto \{0,1,1\}$, $2 \mapsto \{1,0,1\}$, $3 \mapsto \{1,1,0\}$.

It is also possible to use few separate {\bf P-measurements}. 
Scheme of all outcomes for consequent measurements $\op J_x^2$, $\op J_y^2$,
$\op J_z^2$ is depicted below (Tab.~\ref{TabMeasJ}) and it can be found, that 
$p_{\{0,1,1\}} = |k_1|^2$, 
$p_{\{1,0,1\}} = (1-|k_1|^2)\frac{|k_2|^2}{1-|k_1|^2} = |k_2|^2$,
$p_{\{1,1,0\}} = (1-|k_1|^2)\frac{|k_3|^2}{1-|k_1|^2} = |k_3|^2$.
It is also possible to write probabilities of measurement of values 0 and 1
for any $\op J_\nu^2$, {\em e.g.} $p(J_x^2=0)= |k_1|^2$, 
\mbox{$p(J_x^2=1)= 1-|k_1|^2$},
$p(J_y^2=0)= |k_2|^2$, $p(J_y^2=1)= 1-|k_2|^2$, 
$p(J_z^2=0)= |k_3|^2$, $p(J_z^2=1)= 1-|k_3|^2$.

\begin{table}[htbp]
\begin{equation*}
\setlength{\arraycolsep}{2pt}
\newcommand{\arst}[1]{\stackrel{\textstyle #1}{\longrightarrow}}
\newcommand{\toparst}[1]{\stack{\\\\\stackrel{\textstyle #1}{\nearrow}}}
\newcommand{\botarst}[1]{\stackrel{\textstyle #1}{\searrow}}
\begin{array}{c|c|c|c|c|c|c}
     &\op J_x^2&            &\op J_y^2&           &\op J_z^2& \\\hline
     & \toparst{0}&p=|k_1|^2;\ \Mat{c}({1\\0\\0})
     & \arst{1}&p=1;\ \Mat{c}({1\\0\\0})
     & \arst{1}&\stack{p=1;\\\Mat{c}({1\\0\\0})}\\ 
  &&\hrulefill&\hrulefill&\hrulefill&\hrulefill&\hrulefill\\
\Mat{c}({k_1\\k_2\\k_3})& &
      & \toparst{0}
      &\stack[l]{p=\frac{|k_2|^2}{1-|k_1|^2};\\\Mat{c}({0\\1\\0})}
      & \arst{1}&\stack{p=1;\\\Mat{c}({0\\1\\0})}\\
     & \botarst{1}&\smash{\stack[l]{p=|k_2|^2+|k_3|^2\\~\: =1-|k_1|^2;\\
         \dfrac{1}{\sqrt{1-|k_1|^2}}\Mat{c}({0\\k_2\\k_3})}} 
 &&\hrulefill&\hrulefill&\hrulefill\\
&&    
 & \botarst{1}
 &\stack[l]{p=\frac{|k_3|^2}{1-|k_1|^2};\\\Mat{c}({0\\0\\1})}
      & \arst{0}&\stack{p=1;\\\Mat{c}({0\\0\\1})}
\end{array}
\end{equation*}
\caption{Scheme of measurement of $\op J_x^2$, $\op J_y^2$, $\op J_z^2$.}
\label{TabMeasJ}
\end{table}

For any order of measurements the probabilities are the same. Using
more general arguments, it is possible to say, that for commuting
operators different procedures of {\bf P-measurements} simply reproduce
classical statistical manipulations with $\sigma$-algebra with three 
elements and measure $\{|k_1|^2,|k_2|^2,|k_3|^2\}$.

It is also possible to consider some other frame with the same 
$\op J_x^2$ and other $\op J_{y'}^2$, $\op J_{z'}^2$. It corresponds
to decomposition of a state using other $\op\K'$ with frame matrix 
$\op R_{\theta}$ \Eq{Rxy} and eigenvectors \Eq{EK'} 
$\ket{k'_1}=\ket{k_1}$, $\ket{k'_2}$, $\ket{k'_3}$, 
{\ie} with only two changed projectors and probabilities 
$\{|k_1|^2,|k_2'|^2,|k_3'|^2\}$, so here again 
$p(J_x^2=0)= |k_1|^2$ and $p(J_x^2=1)= 1-|k_1|^2$.

So statistical model should not be treated as {\em contextual}, but 
modeling with deterministic computer and random number generator may
essentially depend on order of operations.

The Kochen-Specker theorem disproves possibility of specific hidden 
variables model, that could provide {\em more detailed deterministic 
description of system}, than quantum mechanics itself or minimal model 
from \Sec{SimQC}. Even if using some analogues of Kochen-Specker model, 
like design with slightly nonorthogonal directions discussed in 
\cite{FinKS,BK4} and criticized in \cite{Per3,App3}, we could produce
some deterministic description, it would be quite contradictory achievement, 
because let us talk about things impossible in usual quantum mechanics, like 
simultaneous values for noncommuting operators\footnote{Yet, there are
some arguments in support of {\em the counterfactual reasoning} in 
\cite{P}.}.

\section{Bell pair}
\label{Sec:NlBell}
\subsection{Nonlocality and relativity}
\label{Sec:IntroBell}

The problem with deterministic model is especially clear, when
order of operations is not determined, {\em e.g.}, due to relativity
principle it may be not defined, whose measurement was first,
or second, or it was all simultaneous. So statistical model may
be invariant, because probabilities are not depend on order, but
deterministic implementation has the problems with principle of 
relativity.

Model described in \Sec{SimQC} become nonlocal, then requires possibility 
of instanteneous operations on whole system with few separate parts.
There is standard way to show problems with such description, that
was introduced first in famous Einstein-Podolsky-Rosen paper \cite{EPR}
and discussed further by Bell \cite{BellEPR,BellHidd}. 

The simultaneous measurement procedure may be described 
via set of projectors $\Ps_j \otimes \Ps_k$. Another method --- is to
use ``partial'' measurements, {\ie} separate procedure for
first or second system, described with $\Ps_j \otimes \Id$
or $\Id \otimes \Ps_k$ respectively. It is possible also
to chose different bases $\Ps_k'$ for each system and to use
other commuting sequences of measurements like $\Ps_j \otimes \Ps_k'$,
$\Ps_j \otimes \Id$, $\Id \otimes \Ps_k'$.
 
\smallskip

Let us consider so-called Bell state of two spin systems
\begin{equation}
\ket{B} =  \frac{\sqrt{2}}{2}\bigl(\ketup\ketdn-\ketdn\ketup\bigr)
= \frac{\sqrt{2}}{2}\bigl(\ket{0}\ket{1}-\ket{1}\ket{0}\bigr).
\label{BeSt}
\end{equation}
Here second expression is used in theory of quantum computation,
there system with two states is called {\em qubit} with notation 
$\ket{0}$ and $\ket{1}$ for two basic states.

The {\em singlet} state \Eq{BeSt} does not depend on change of basis
\begin{equation}
\ket{0'} = \alpha \ket{0} + \beta \ket{1}, \quad
\ket{1'} = \bar\alpha \ket{1}-\bar\beta \ket{0}, \quad
\ket{0'}\ket{1'}-\ket{1'}\ket{0'} = \ket{0}\ket{1}-\ket{1}\ket{0}.
\label{indep}
\end{equation}

\subsection{Description of single system (qubit)}
\label{Sec:Qubit}

It is convenient to consider first property of single spin system or qubit.
The spin operators $\op S_\nu = \frac{1}{2}\op \sigma_\nu$ ($\nu = x,y,z$) 
are expressed via Pauli matrixes 
\begin{equation}
\op \sigma_x = \Mat{cc}({0&1\\1&0}), \quad
\op \sigma_y = \Mat{cc}({0&-i\\i&0}), \quad
\op \sigma_z = \Mat{cc}({1&0\\0&-1}).
\label{Pauli}
\end{equation}

It is possible also to introduce set with three pairs of
orthogonal projectors 
\begin{equation}
\Ps_\nu^{\pm} = \frac{1}{2}(\op \bbId \pm \op\sigma_\nu),\quad
\Ps^+_\nu + \Ps^-_\nu = \op \bbId, \quad
\nu = x,y,z.
\label{Pxyz}
\end{equation}
They corresponds to measurement of spin $\pm 1/2$ along axes
$x,y,z$.

Measurement along arbitrary axis $\ve{a}=(a_x,a_y,a_z)$ is
described using projectors
\begin{equation}
 \Ps_\ve{a} = \frac{1}{2}(\op \bbId + 
  a_x \op\sigma_x + a_y \op\sigma_y + a_z \op\sigma_z),\quad
  a_x^2 + a_y^2+a_z^2 = 1, 
\label{Pa}
\end{equation}
where two orthogonal projectors correspond to \Eq{Pa} with
two opposite vectors $\pm\ve{a}$. It justifies geometrical
model (Bloch sphere) with directions in 3D space for spin-half 
system, but the same algebraic formalism may be used for 
any quantum system with two-dimensional Hilbert space,
{\ie} {\em qubit}.

Let us write relations between vectors $\pm\ve{a}$ used for
construction of projectors \Eq{Pa} and states of qubit
\begin{equation}
 \ket{\psi} = \alpha \ket{0} + \beta \ket{1},\quad 
 |\alpha|^2 + |\beta|^2 = 1
\label{psi}
\end{equation}
The projector is represented as $\Ps = \ket{\psi}\bra{\psi}$ and
due to \Eq{Pa}
$$
\ket{\psi}\bra{\psi} =
 \Mat{cc}({\alpha\bar\alpha & \alpha\bar\beta \\
 \beta\bar\alpha & \beta\bar\beta}) =
 \frac{1}{2}\Mat{rl}({1 + a_z & a_x - i a_y \\  a_x + i a_y & 1 - a_z}),
$$
so, it is possible to find all components of vector $\ve{a}$
\begin{equation}
 a_x = \alpha\bar\beta + \bar\alpha\beta,\quad
 a_y = i(\alpha\bar\beta - \bar\alpha\beta),\quad
 a_z = \alpha\bar\alpha - \beta\bar\beta.
\label{q2v}
\end{equation}
It is clear, that \Eq{q2v} do not depend on phase and so represent
correct map from space of states, {\ie} rays in Hilbert space, to 
sphere $\Sph^2$ of unit vectors in $\R^3$.

\begin{figure}[hbt]
\begin{center}
\includegraphics[scale=0.5]{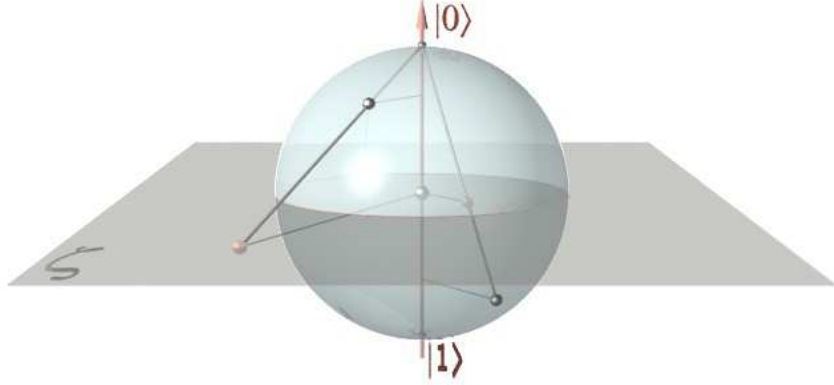}
\end{center}
\caption{Stereographic projection from Bloch sphere}\label{Fig:bloch}
\end{figure}

The opposite map is more complicated due to phase ambiguity 
of qubit \Eq{psi}. One visual method to avoid that --- is to use
{\em ``Argand plane with point at infinity''}: $\zeta = \alpha/\beta$ 
\cite{PenRinI}. Such representation maps $\ket{1}$ to origin of 
the plane and $\ket{0}$ to the point at infinity. The map of the 
Bloch sphere to the plane is {\em stereographic projection} (\Fig{bloch})
\begin{equation}
\zeta = (a_x - i a_y)/(1-a_z)
\label{ster}
\end{equation}

Inverse map may be written as
\begin{equation}
a_x = \frac{2 \RE \zeta}{|\zeta|^2+1},\quad
a_y = \frac{-2 \IM \zeta}{|\zeta|^2+1},\quad
a_z = \frac{|\zeta|^2-1}{|\zeta|^2+1}.
\label{rest}
\end{equation}

For concrete calculations it is better to work with usual
pair $\alpha\ket{0}+\beta\ket{1}$ \Eq{psi} 
instead of \Eq{ster} with $\infty$ for $a_z=1$. It is not possible 
to describe that using one map, {\ie}
such inverse construction for \Eq{q2v} may be expressed
as {\em atlas} with two hemispheres
\begin{equation}
\ket{\psi_\ve{a}} =
\begin{cases}
 \displaystyle
 \frac{1}{\sqrt{2(1+a_z)}}
  \bigl((1+a_z)\ket{0}+(a_x + i a_y)\ket{1}\bigr); 
 & a_z \ge 0,\\ 
 \displaystyle
  \frac{1}{\sqrt{2(1-a_z)}}
 \bigl((a_x - i a_y)\ket{0}+(1-a_z)\ket{1}\bigr); 
 & a_z < 0.
\end{cases}
\label{v2q}
\end{equation}
Both maps in atlas presented in \Eq{v2q} defined everywhere, 
except one pole of sphere $a_z=-1$ and $a_z=1$, respectively. 
For such definition on sphere without poles both maps \Eq{v2q}
overlap, but differ on a phase multiplier 
$$(a_x + i a_y)/\sqrt{1-a_z^2}=(a_x + i a_y)/\sqrt{a_x^2+a_y^2}$$
and so represent the same quantum state. 

It is possible to use division of sphere using other 
value $a_z$ as threshold in \Eq{v2q}, say in degenerate case $a_z = -1$ second 
hemisphere contracts to single point on pole. But it 
is impossible to find a single {\em continuous} map due 
to some general theorem of {\em differential geometry}\footnote{
It was already mentioned, that the space of normalized complex 
vectors like \Eq{psi} is equivalent to (hyper)sphere. For qubit
it is $\Sph^3$ and so \Eq{q2v} correspond to map 
$\Sph^3 \to \Sph^2$. The map is well known as Hopf fibration
and absence of continuous inverse map (section) follows from
general theory of fiber bundles \cite{fibr}.}. 

It is convenient to consider a case with projectors ``perpendicular to
$z$ axis,'' {\ie} equator of Bloch sphere with $a_z=0$ and so due to \Eq{v2q}
\begin{equation}
 \ket{\phi_\ve{a}} = \frac{1}{\sqrt{2}}{\bigl(\ket{0}+(a_x+i a_y)\ket{1}\bigr)}. 
\label{v2q2}
\end{equation}

It should be mentioned also, that density matrix of any pure
state \Eq{rho} does not have any problems with phase ambiguity and 
defined by obvious analogue of \Eq{Pa}
\begin{equation}
  \rho_{\ve{a}} = \ket{\psi_{\ve{a}}}\bra{\psi_{\ve{a}}} = 
  \frac{1}{2}(\op \bbId + a_x \op\sigma_x + a_y \op\sigma_y + a_z \op\sigma_z).
\label{ra}
\end{equation}
Probabilities of measurements may be represented due to
\Eq{prho} as
\begin{equation}
  \Tr(\Ps_{\ve{b}} \op\rho_{\ve{a}})
 = \tfrac{1}{2}(1 + a_x b_x + a_y b_y + a_z b_z) =
 \tfrac{1}{2}+\tfrac{1}{2}\scprod{\ve{a}}{\ve{b}}. 
\label{pab}
\end{equation}

\subsection{Measurement of Bell pair}
\label{Sec:MeasBell}

Let us now consider measurements of two qubits in singlet
state \Eq{BeSt}. The principle resembles different 
procedures with P-measurement in \Sec{KSphys}.

Let us for simplicity to consider for both systems only 
measurements described by projectors \Eq{Pa} with $a_z=0$ 
and states \Eq{v2q2}. For such a kind of states it is
simple to calculate scalar products with Bell state \Eq{BeSt}
\begin{equation*}
 B_{a,b} =
 \frac{\sqrt{2}}{2}
 \bigl(\bra{0}\bra{1}-\bra{1}\bra{0}\bigr)\ket{\phi_\ve{a}}\ket{\phi_\ve{b}} =
 \frac{\sqrt{2}}{4}(b_x + i b_y - a_x - i a_y)
\end{equation*}
and standard expression for probabilities\footnote{The expression 
with $\scprod{\ve{a}}{\ve{b}}$ is valid for arbitrary states 
(without limitation $a_z=b_z=0$).}
\begin{equation}
 p_{\ve{a},\ve{b}} = |B_{a,b}|^2 = 
 \frac{1}{4}\bigl(1 - \scprod{\ve{a}}{\ve{b}}\bigr).
\label{pBe}
\end{equation}

Any pair of states $\ket{\phi_\ve{a}}$ and $\ket{\phi_{-\ve{a}}}$ is
the basis in two-dimensional Hilbert space. Four projectors 
$\Ps_{\pm\ve{a}} \otimes \Ps_{\pm\ve{b}}$ describe measurement 
procedure for system with two parts. Probabilities
of four different outcomes $\ket{\phi_{\pm\ve{a}}}\ket{\phi_{\pm\ve{b}}}$
are described by \Eq{pBe}, {\ie}
$\{p_{\ve{a},\ve{b}},p_{-\ve{a},\ve{b}},p_{\ve{a},-\ve{b}},p_{-\ve{a},-\ve{b}}\}$  
or
\begin{equation}
 \left\{
 \frac{1}{4}\bigl(1 - \scprod{\ve{a}}{\ve{b}}\bigr),
 \frac{1}{4}\bigl(1 + \scprod{\ve{a}}{\ve{b}}\bigr),
 \frac{1}{4}\bigl(1 + \scprod{\ve{a}}{\ve{b}}\bigr),
 \frac{1}{4}\bigl(1 - \scprod{\ve{a}}{\ve{b}}\bigr)
 \right\}.
\label{allBe}
\end{equation}
Projectors $\Ps_L=\Ps_{\pm\ve{a}} \otimes \Id$ and
$\Ps_R=\Id \otimes \Ps_{\pm\ve{b}}$ describe ``partial'' measurements on first 
and second subsystem respectively. Here situation is similar
with measurement of $\op J_\nu^2$ components, because the
operators also have degenerate eigenspaces. 

Formally \Eq{allBe} show, that probabilities to find subsystems
in two different states are equal 
$p_{\ve{a},*} \equiv p_{\ve{a},\ve{b}}+p_{\ve{a},-\ve{b}} = 1/2 = p_{-\ve{a},*}$,
and the same with second one
$p_{*,\ve{b}} \equiv p_{\ve{a},\ve{b}}+p_{-\ve{a},\ve{b}}=1/2=p_{*,-\ve{b}}$.
So results of measurements are correlated
$p_{\ve{a},\ve{b}} = (1 - \scprod{\ve{a}}{\ve{b}})/4
\ne p_{\ve{a},*}\, p_{*,\ve{b}}=1/4$.
It is usually considered {\em correlator}
\begin{equation}
 E(\ve{a},\ve{b}) =
 p_{\ve{a},\ve{b}}-p_{-\ve{a},\ve{b}}-p_{\ve{a},-\ve{b}}+p_{-\ve{a},-\ve{b}}
 = - \scprod{\ve{a}}{\ve{b}}.
\label{corBe}
\end{equation}

The \Eq{corBe} describes  ``expectation value of the product of the 
two components,'' \cite{BellHidd} {\ie} two components with values $\pm 1$
for measurement of spin along or opposite of given axes\footnote{It is
denoted as $P({\ve{a},\ve{b}})$ in \cite{Bell,BellHidd} and as $\langle ab \rangle$
in \cite{P}.}.

It is possible also to use procedure of {\bf P-measurement} 
from \Sec{SimQC} with application of $\Ps_L$ to Bell state $\ket{B}$. 
It produces $\ket{\phi_{\ve{a}}}\ket{\phi_{-\ve{a}}}$ or 
$\ket{\phi_{-\ve{a}}}\ket{\phi_{\ve{a}}}$ with equal 
probabilities $1/2$. If it to apply $\Ps_R$ measurement after that, 
the final result of two such measurements
is one of four states $\ket{\phi_{\pm\ve{a}}}\ket{\phi_{\pm\ve{b}}}$
and \Eq{pab} ensures, that probabilities are the
same \Eq{allBe} as for simultaneous measurement of
all values described earlier. 
The result is the same also if $\Ps_R$ measurement is applied
before $\Ps_L$. 

So statistical results of the model do not depend on order of measurements
and do not contradict relativity and locality, but simulation
that on computer with random number generator has problems with
the principles. 

\smallskip

The \Eq{corBe} coincides with quantum mechanical
expectation value and so any combinations with different orientations
are {\em in agreement with quantum mechanical description}.
{\em E.g.}, Bell inequality in initial form \cite{P,BellHidd}
\begin{equation}
 |E(\ve{a},\ve{b}) - E(\ve{a},\ve{c})| \le 1 - E(\ve{b},\ve{c}),
\label{ineqBell}
\end{equation}
or more general version, CHSH (Clauser-Horne-Shimony-Holt)
inequality \cite{P,CHSH}
\begin{equation}
 |E(\ve{a},\ve{b}) + E(\ve{b},\ve{c}) + 
 E(\ve{c},\ve{d}) - E(\ve{d},\ve{a}) | \le 2
\label{ineqCHSH}
\end{equation}
must be {\em violated} for some directions. 

\smallskip

Such agreement with quantum mechanical description is produced
by formal nonlocality. Really already description of
state by single array of complex numbers with four component
used in model \Sec{SimQC} is hardly may be treated as local.

\begin{figure}[hbt]
\begin{center}
\includegraphics[scale=0.5]{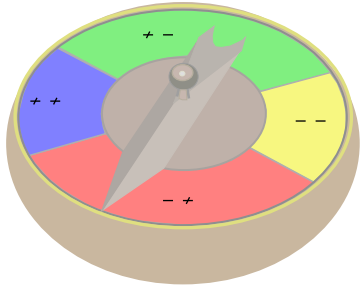}
\end{center}
\caption{Roulette used for nonlocal modeling of Bell pair}\label{Fig:rulBell}
\end{figure}

Model of measurement procedure is also not local,
for example for description of simultaneous measurement
of both systems should be used single random number for generating
of four probabilities $\{p_{++},p_{+-},p_{-+},p_{--}\}$ \Eq{allBe}. 
It is similar with single roulette (\Fig{rulBell}) with four sectors 
(with angles $2\pi p_{++}$, $2\pi p_{+-}$, $2\pi p_{-+}$, $2\pi p_{--}$), 
when given sector corresponds to appearance of pair of values for
two distant measurements.

On the other hand, simulation of classical correlations
also may use similar description, despite of locality
of simulated physical system. But for classical case local
description also exists.

The nonlocality is more obvious for separate measurements,
then model \Sec{SimQC} used above produces two outcomes like
$\ket{\phi_{\ve{a}}}\ket{\phi_{-\ve{a}}}$ and 
$\ket{\phi_{-\ve{a}}}\ket{\phi_{\ve{a}}}$ 
after ``partial'' measurement. In computer simulation
it is some intermediate result, hidden from a user
of the program, but it just coincides with description of
``instantaneous reduction,'' criticized in EPR paper \cite{EPR}.

\section{Conway-Kochen ``Free Will'' theorem}
\label{Sec:3W}

If to use two separate spin-1 systems, then combination
of Kochen-Specker theorem with nonlocality complicates 
the picture. 

Here is considered setup with more detailed physical model  
used in \cite{KC}. For each system is considered operators
$\op J_\nu$ \Eq{JsP} and $\op J_\nu^2$ \Eq{EJP}. Here they
are written in symmetric form. Let us find precise expression 
for singlet state of two spin-1 systems with respect to such 
operators.

The singlet state should be invariant with respect to rotation
of sphere in 3D Euclidean space used in physical model of
Kochen-Specker theorem, {\ie} `$\op\K$-sphere' described in \Sec{KSphys}.
It may be simply described using symmetric composition with two bases of 
the 3D real space \Eq{EKN}
\begin{equation}
\ket{\TW}
 =\frac{1}{\sqrt{3}}\bigl(\ket{1_\K}\ket{1_\K}+\ket{2_\K}\ket{2_\K}
 +\ket{3_\K}\ket{3_\K}\bigr),
\label{Ksngl}
\end{equation}
because it is invariant with respect to Euclidean 3D rotations.
It can be checked directly, that
for any $\nu = x,y,z$ \Eq{Ksngl} may be rewritten using \Eq{EJP} as
$$
\frac{1}{\sqrt{3}}\Bigl(\ket{0_\nu}\ket{0_\nu}
 +\frac{1}{2}(\ket{+_\nu}{+}\ket{-_\nu})(\ket{+_\nu}{+}\ket{-_\nu})
 +\frac{i^2}{2}(\ket{+_\nu}{-}\ket{-_\nu})(\ket{+_\nu}{-}\ket{-_\nu})\Bigr).
$$
and so expressed via eigenvectors \Eq{EJP} of any $\op J_\nu$
\begin{equation}
\ket{\TW}=\frac{1}{\sqrt{3}}\bigl(\ket{0_\nu}\ket{0_\nu}+
 \ket{+_\nu}\ket{-_\nu}+\ket{-_\nu}\ket{+_\nu}\bigr)
\label{Jsngl}
\end{equation}

It should be mentioned, that \Eq{Ksngl} and \Eq{Jsngl} are invariant only
with respect to SO(3) subgroup of SU(3) group\footnote{
The difference of sign in first term of \Eq{Jsngl} with \cite{KC} is
consequence of this fact.}.

\medskip

Let us apply simultaneous {\bf N-measurement} (here $N= 3 \times 3 = 9$)
of state $\ket{\TW}$ \Eq{Jsngl} with two different frames defined by 
eigenvectors of $\op\K$ \Eq{EKN} and $\op\K'$ \Eq{EK'}. Using representation of 
$\ket{\TW}$ \Eq{Ksngl}, and equation 
\mbox{$\ket{i_\K}=\sum_{j=1}^3 R_{ji}\ket{j_\K'}$}, it is possible
to find decomposition for any basis $\ket{i_\K}\ket{j_\K'}$
\begin{equation}
 \ket{\TW} = \frac{\sqrt{3}}{3}\sum_{i,j=1}^3 R_{ji}\ket{i_\K}\ket{j_\K'},
\label{TWij}
\end{equation}
there $R_{ji}$ are elements of transposed matrix \Eq{Rot}. So the
measurement produces up to nine outcomes $\ket{i_\K}\ket{j_\K'}$ with
probabilities $\frac{1}{3}|R_{ji}|^2$, but for description of
experiments discussed in \cite{KC} such a fine decomposition
is not required.
 
In \cite{KC} is considered case, when first experimenter measures 
full frame $(\ve{x},\ve{y},\ve{z})$ and second one only one
direction. Let it be measurement of $\op J_x^2$. 
It is possible to apply {\bf $\op\K \otimes \op J_x^2$-measurement}
or {\bf P-measurement} procedure from
\Sec{SimQC} for such scenario with $3\times 2 = 6$ projectors.

The results for both experimenters 
$\bigl\{\underbrace{\K \Rightarrow \{J_x^2,J_y^2,J_z^2\}}_{\rm I}, 
  \underbrace{J_x^2}_{\rm II}\bigr\}$ are:
\begin{eqnarray*}
&&\bigl\{1 \Rightarrow \{0,1,1\},~~0\bigr\}, ~ p=1/3,~ \text{state } 
 \ket{1_\K}\ket{1_\K} = \ket{1_\K}\ket{0_x}, \\
&&\bigl\{2 \Rightarrow \{1,0,1\},~~1\bigr\}, ~ p=1/3,~ \text{state }
 \ket{2_\K}\ket{2_\K} = \ket{2_\K}\frac{i}{\sqrt{2}}(\ket{+_x}-\ket{-_x}), \\ 
&&\bigl\{3 \Rightarrow \{1,1,0\},~~1\bigr\}, ~ p=1/3,~ \text{state }
 \ket{3_\K}\ket{3_\K} = \ket{3_\K}\frac{1}{\sqrt{2}}(\ket{+_x}+\ket{-_x}),\\
\end{eqnarray*}
{\ie} results for $J_x^2$ always in agreement for both sides. 
The similar is true for measurement $\op J_y^2$ and $\op J_z^2$, {\ie}
both experimenters got the same results for given axis.

Here is only three elements $\ket{j_\K}\ket{j_\K}$ instead of 
nine in \Eq{TWij}, because 
$\op\K$ and $\op J_x^2$ commute and basis $\ket{j_\K}$ may
be used for decomposition for both sides. It is not necessary
even to know other elements of frame $\op J_{y'}^2$ and
$\op J_{z'}^2$ for second experimenter to predict results
of measurement.

Model of \Sec{SimQC} may ensure such agreement with quantum mechanics due
to `nonlocal roulette' generating value of both measurements and alredy 
discussed in \Sec{MeasBell}. So, example with two spin-1 systems 
discussed here does not differ much from consideration of Bell pair,
or maybe even simpler, if to use model defined in \Sec{SimQC}.

\smallskip

 In the main, model with two spins discussed in Conway-Kochen ``Free Will''
theorem \cite{KC} and some works on ``quantum pseudotelepahy'' \cite{QPT}
is interesting due to comparison with particular classical 
description. Really, it is possible to suggest a classical model that produces 
agreement, if for different frames second experimenter is using $x$. 
Say, it is enough to use 
two equal programs on two distant computers with procedure \Eq{gen3}
described in \Sec{KSimp}. With same value $r$ of pseudo-random number
generator both users will have the same results for the frames with 
the same $x$. On the other hand, Kochen-Specker theorem ensures, 
that such a method {\em fails}, if second user may choose also 
$y$ or $z$ axis, {\em e.g.,} to use second or third outcome in \Eq{gen3}. 

The problem is similar with contextuality due to hidden dependence 
$p_y$ in algorithms like \Eq{gen3} from value $p_x$ already discussed 
in \Sec{KSimp}. Using two equivalent 
algorithms with pseudo-random numbers let reveal that, and Kochen-Specker 
theorem closes any possibility to escape\footnote{But formally, {\em it is only
no-go result for hidden variables models of the same classes, which are used in 
Bell, Kochen-Specker and Conway-Kochen models}. It may be risky to be 
absolutely sure, that any local classical model will not work, because history 
of no-go theorems about hidden variables have shown the appearance of new and new 
models, which were not taken into account by authors of no-go theorems.}. So from 
{\em classical point} of view behavior of the quantum mechanical system 
(that anyway always produces the same results) is really 
miracle. 
Otherwise (without reference to the classical model) it is very similar with
nonlocality problems associated with Bell pair and already discussed earlier
in relation with ``free will'' \cite{PerW}.

\section{Modified Deutsch-Hayden-Tipler model}
\label{Sec:DHT}

Some conceptual problems with model in \Sec{SimQC} were already mentioned.
It maybe even possible to accept hidden internal nonlocality, that
may not be found by any statistical test, but problem with relativity
of order of events make the model nonconstructive. 
 
There is interesting method to avoid problem with nonlocality and relativity
discussed in \cite{DeuHay,Tip0}. The papers use interpretation of quantum
mechanics without reduction \cite{Ev}, but may be adapted for
interpretation with ``local reduction condition'' to meet
relativity principle. The basic idea, is that
any measurement of correlation is possible only after final, local
measurement, when some carriers with data about two nonlocal measurements 
in distant points are reunited.

\begin{figure}[hbt]
\begin{center}
\parbox[c]{5cm}{
\includegraphics[scale=0.4]{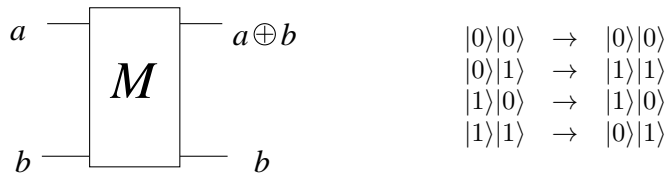}
}
\parbox[c]{4cm}{\hfill
$
\begin{array}{lcr}
 \ket{0}\ket{0} & \to & \ket{0}\ket{0} \\
 \ket{0}\ket{1} & \to & \ket{1}\ket{1} \\
 \ket{1}\ket{0} & \to & \ket{1}\ket{0} \\
 \ket{1}\ket{1} & \to & \ket{0}\ket{1} 
\end{array}
$
}
\end{center}
\caption{Non-perturbing measurement gate}\label{Fig:qmeas}
\end{figure}

It may be simply described using model \Sec{SimQC}. The basic
tool for two distant measurement is {\em non-perturbing measurement
gate} \cite{DeuQN,DeuHay} \Fig{qmeas}.

Really, in \cite{DeuHay} was used Heisenberg representation
of quantum computations \cite{GotHeis}. Such representation
has some advantages and formally was used already in \cite{FeySim},
but it should be discussed elsewhere and here is used Schr\"odinger
picture adopted in model \Sec{SimQC}.

The measurement gate is described by formula 
\begin{equation}
\op M : \ket{a}\ket{b} \to \ket{a \oplus b}\ket{b},
\quad \op M = \Id \otimes \Ps_0 + \op\sigma_x \otimes \Ps_1,
\label{MeasGate}
\end{equation}
where $\oplus$ is addition modulo two. For $a=0$ such
gate performs formal measurement in chosen (`compuational')
basis, {\ie} copies second state $\ket{b}$ to the first one. 
Here note about {\em computational basis} is essential, because it 
is impossible to clone arbitrary state of $\ket{b}$ \cite{NoClon}.

Let us consider measurement of Bell pair. 
$$ \frac{1}{\sqrt{2}}\bigl(\ket{0}\ket{1} - \ket{1}\ket{0}\bigr). $$
At first step to each part ($A$ and $B$) of distant pair are attached 
auxiliary qubits:
$$ \ket{0_A}\Bigl(
   \frac{1}{\sqrt{2}}\bigl(\ket{0}\ket{1} - \ket{1}\ket{0}\bigr)
   \Bigr)\ket{0_B}. $$
After application of non-destructive measurement gates \Eq{MeasGate} 
$\op M$ the state is
$$ \frac{1}{\sqrt{2}}\bigl(\ket{0_A}\ket{0}\ket{1}\ket{1_B} - 
   \ket{1_A}\ket{1}\ket{0}\ket{0_B}\bigr).$$
It is now possible to reunit auxiliary qubits in some local place
$$ \frac{1}{\sqrt{2}}\bigl(\ket{0_A}\ket{1_B}\,\ket{0}\ket{1} - 
   \ket{1_A}\ket{0_B}\,\ket{1}\ket{0}\bigr)$$
and perform destructive measurement that produces
either $\ket{0_A}\ket{1_B}\,\ket{0}\ket{1}$ or
$\ket{1_A}\ket{0_B}\,\ket{1}\ket{0}$ with equal probabilities.

From the one hand, such a picture does not have the problem with relativity 
of notion of simultaneous events discussed earlier, because qubits $A$ and $B$ 
are not separated. From the other one, presented elementary model hardly could 
be considered as local, because state of whole nonlocal system with four 
qubits after final measurement is changed instantly. Similar models in
\cite{DeuHay,Tip0} are local, because any destructive measurements 
(`collapse of wave function') are forbidden.

\begin{figure}[hbt]
\begin{center}
\parbox[c]{5cm}{
\includegraphics[scale=0.4]{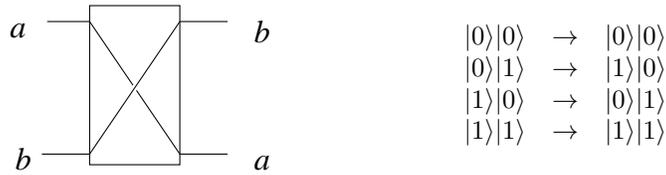}
}
\parbox[c]{4cm}{\hfill
$
\begin{array}{lcr}
 \ket{0}\ket{0} & \to & \ket{0}\ket{0} \\
 \ket{0}\ket{1} & \to & \ket{1}\ket{0} \\
 \ket{1}\ket{0} & \to & \ket{0}\ket{1} \\
 \ket{1}\ket{1} & \to & \ket{1}\ket{1} 
\end{array}
$
}
\end{center}
\caption{Swap (exchange) gate}\label{Fig:qswap}
\end{figure}

A simple way to avoid the last problem with nonlocality, is to use
instead of measurement gate \cite{DeuHay,Tip0} {\sf swap} (exchange)
gate (\Fig{qswap})
\begin{equation}
\op E : \ket{a}\ket{b} \to \ket{b}\ket{a},
\quad \op E = \frac{1}{2}( \Id \otimes \Id +  \op\sigma_x \otimes \op\sigma_x 
+ \op\sigma_y \otimes \op\sigma_y +  \op\sigma_z \otimes \op\sigma_z).
\label{SwapGate}
\end{equation}

This gate exchanges any two states, not only basic ones, and
not affected by {\em no-cloning} theorem. With application of
such a gate the procedure of measurement discussed above becomes
simpler:
 
{\em First step} 
$$ \ket{0_A}\Bigl(
   \frac{1}{\sqrt{2}}\bigl(\ket{0}\ket{1} - \ket{1}\ket{0}\bigr)
   \Bigr)\ket{0_B}. $$

{\em Second step} (two distant `swap'-measurements)
$$ \frac{1}{\sqrt{2}}\bigl(\ket{0_A}\ket{0}\ket{0}\ket{1_B} - 
   \ket{1_A}\ket{0}\ket{0}\ket{0_B}\bigr).$$

{\em Third step} (reunion of $A$ and $B$)
$$ \frac{1}{\sqrt{2}}\bigr(\ket{0_A}\ket{1_B}
 -\ket{1_A}\ket{0_B}\bigl)\,\ket{0}\ket{0}.$$

So even final destructive measurement, that produces with equal probabilities
either $\ket{0_A}\ket{1_B}\,\ket{0}\ket{0}$ or
$\ket{1_A}\ket{0_B}\,\ket{0}\ket{0}$, does not affect on
states $\ket{0}$ of distant pair of qubits.

\section*{Acknowledgments and disclaimers}

Author is grateful to participants of {\em A. Friedmann Lab} seminar  
(8 June 2006, St.-Petersburg, Russia) for comments and attention.
This work is not made for hire and is not funded from any sources.

\end{document}